\documentclass[prb,
showpacs,floatfix,prb
]{revtex4-1}
\usepackage{latexsym}
\usepackage{amssymb}
\usepackage{amsfonts}
\usepackage[usenames]{color}
\usepackage{epsfig}
\newcommand{\vecrep}{\mathbf}
\newcommand{\area}{\mathcal{A}}
\newcommand{\KKp}{$K\!-\!K^{\prime}$}
\newcommand{\AB}{$A-B$}

\newcommand{\jj}{{\overline\jmath}}

\newcommand{\ket}[1]{\left| #1 \right>}
\newcommand{\bra}[1]{\left< #1 \right|}

\begin{document}
\title{Coherent transport through graphene nanoribbons in the presence of edge disorder}
\author{F Libisch$^{1,2}$, S Rotter$^1$ and J Burgd\"orfer$^1$}
\address{$^1$Institute for Theoretical Physics, Vienna University of 
Technology\\Wiedner Hauptstra\ss e 8-10/136, A-1040 Vienna, Austria, European
Union}
\address{$^2$Department for Mechanical and Aerospace Engineering, Princeton
University, Olden Str., Princeton, NJ 08544, USA}
\date{\today}

\begin{abstract}
We simulate electron transport through graphene nanoribbons of
experimentally realizable size (length $L$ up to $2\mu$m, width
$W\approx 40$nm) in the presence of scattering at rough edges. Our
numerical approach is based on a modular recursive Green's function
technique that features sub-linear scaling of the computational effort
with $L$. We investigate backscattering at edge defects: Fourier
spectroscopy of individual scattering states allows us to disentangle
inter-valley and intra-valley scattering. We observe Anderson
localization with a well-defined exponential decay over 10 orders of
magnitude in amplitude. We determine the corresponding localization
length for different strength and shape of edge roughness.
\end{abstract}



\maketitle

\section{Introduction}

\begin{figure}[t]
\hbox{}\hfill\epsfig{file=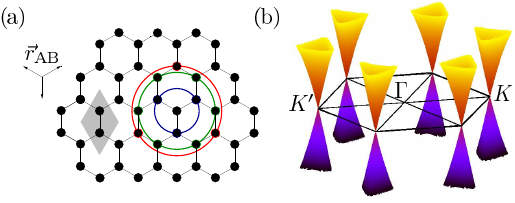, width = 8cm}\hfill\hbox{}
\caption{(a) Graphene hexagonal lattice with lattice
  constant $a = 1.4$\AA. The unit cell (shaded area) contains two
  carbon atoms $A$ and $B$ belonging to the two triangular sublattices
  connected to each other by the displacement vector
  $\vec{r}_{\mathrm{AB}}$ [see inset]. Each
  atom has three nearest neighbours [smallest (blue) circle], six
  next-nearest neighbours [medium (green) circle] and three
  second-nearest neighbours [largest (red) circle]. (b) The conical
  dispersion relation with trigonal warping of an infinite graphene
  plane near the $K$ and $K'$ points, as obtained by the third-nearest
  neighbour tight-binding approach.}
\label{fig:TB}
\end{figure}

The experimental realization of graphene, i.e., of a monolayer of
carbon atoms \cite{BerryGeim, MasslessFirsov, ElectricFirsov} has
opened up a rapidly developing field of fundamental and applied
physics. The topology of the planar honeycomb lattice
[figure \ref{fig:TB}(a)] with the resulting peculiar band structure near
the $K$ and $K'$ points [figure \ref{fig:TB}(b), for a review, see
\cite{guinea_review, dassarma_review}] gives rise to many novel and
intriguing physical properties, including the room temperature quantum
Hall effect, minimum conductivity at the Fermi energy as well as
possible applications for spintronics. Recent advances in fabricating
width-modulated graphene nanoribbons helped to overcome intrinsic
difficulties in creating tunnelling barriers and confining electrons in
graphene, where transport is dominated by Klein tunneling-related
phenomena \cite{dom99,kat06}. Graphene quantum dots have been
fabricated and Coulomb blockade \cite{sta08aa,sta08ab,pon08}, quantum
confinement \cite{sch09} and charge detection \cite{gue08} have been
demonstrated.

The electronic properties of the \emph{perfect} honeycomb lattice are
meanwhile theoretically well understood \cite{guinea_review}. However,
in realistic graphene devices finite-size effects and imperfections
play an essential role, especially for transport through confined
structures. The importance of such effects results from the gapless
band structure of graphene which does not allow straightforward
confinement by electrostatic potentials.  Devices thus have to be cut
or etched resulting in rough edges. In turn, properties of the ideal
graphene band structure cannot be invoked when simulating quantum
transport through realistic devices in the presence of randomly shaped
boundaries. Moreover, recent results underline that an incoherent
Boltzmann transport approach, unlike a full quantum-mechanical
calculation, fails to reproduce experimentally observed conductance
signatures for impurity scattering \cite{Klos}. However, the
application of numerical methods for a full quantum mechanical
simulation of graphene ribbons of realistic size constitutes a
considerable challenge. A method of choice is the widely-used
recursive Green's function technique \cite{MacKinnon2} which is
well suited to treat scattering structures such as wires or ribbons
extended in one of their dimensions and usually implemented within the
Landauer-B\"uttiker framework \cite{Ferry} for calculating transport
coefficients. This technique has meanwhile been incorporated
successfully into both an equilibrium
\cite{bara,Lewenkopf,Roche,Heinzel,tra07} and a non-equilibrium
\cite{cres,meta,neop,sviz} (Keldysh) description. Several variants of
this method have been put forward employing specific symmetries of the
system \cite{sols,PRB62RTWTB,Stephan} if applicable, or using recursive
algorithms \cite{kazy,wimm,drou,mama,JoooSuper,JoooSuper2}. In this
article we present an extension of the modular recursive Green's
function method (MRGM)\cite{Stephan} designed to treat graphene
nano-ribbons with edge or bulk disorder. We address disorder
scattering in graphene nanoribbons both on the microscopic level of
specific lattice defects as well as on the macroscopic level of
current measurements for which, as we will demonstrate, the details of
the underlying disorder scattering play a crucial role.

This paper is organized as follows: We briefly review key
properties of the band structure of ``ideal'' infinitely extended
graphene in the absence of disorder in section \ref{sec:ideal}. In
section \ref{sec:tech}, we introduce the application of the MRGM to
finite-size graphene structures which allows us to treat extended
structures efficiently due to the favorable scaling of the numerical
effort with the linear dimensions of the ribbon. Applications to
transport through rough-edged graphene nanoribbons will be presented
in section \ref{sec:rec} followed by a short summary (section
\ref{Sec:sum}).

\section{Tight-binding simulation of graphene band structure}\label{sec:ideal}

The ideal, infinitely extended graphene sheet features a honeycomb
lattice made up of two (A and B) interleaved triangular sublattices.
It can be described in tight-binding (TB) approximation by the
Hamiltonian \cite{Wallace}
\begin{equation}
  H = \sum_{i,s}\ket{\phi_{i,s}}V_i\bra{\phi_{i,s}}-
  \sum_{(i,j),s}\gamma_{i,j}\ket{\phi_{i,s}}\bra{\phi_{j,s}} +
  h.c.\,, \label{H_Graph_TB}
\end{equation}
where the sum $(i,j)$ extends over pairs of lattice sites,
$\ket{\phi_{j,s}}$ is the tight-binding orbital with spin $s$ at
lattice site $j$, $V_i$ is a locally varying potential (onsite
energy), and $\gamma_{i,j}$ is the hopping matrix element between
lattice sites $i$ and $j$. Within our TB approximation, we include
third-nearest-neighbour coupling [see figure \ref{fig:TB}(a)] using
orthogonal tight-binding orbitals. This allows for four free
parameters, namely the site-energy $\varepsilon_0$ and the overlap
integrals $\gamma_i$, $i = 1,2,3$, representing the interaction with
the first, second and third nearest neighbour, respectively. We 
choose the $\gamma_i$ by fitting to ab-initio
calculations, taken from Reich et al.~\cite{TBReich, Grueneis,
  LibPRB08}, arriving at $\gamma_1 = -3.145$, $\gamma_2 = -0.042$, and
$\gamma_3 = -0.35$.

The dispersion near the non-equivalent $K$ and $K'$ points [figure
  \ref{fig:TB}(b)] resulting from the diagonalization of equation
(\ref{H_Graph_TB}) features [for large distances $k$ from
  $K$($K'$)] deviations from a perfect cone reflecting the influence
of the hexagonal lattice. The cone becomes squeezed along the $K-K'$
directions, an effect known as trigonal warping \cite{McCann06b,
  guinea_review}. Near the $K$ point and for small $k$ the band
structure of equation (\ref{H_Graph_TB}) can be approximated (assuming
that $V_i \ll t$) by a conical dispersion relation around the $K$
point \cite{Semenoff},
\begin{equation}\label{Eq:Semenoff}
  E(k + k_K) = E(k_K) + k \partial_k E(k_K) +
  \mathcal{O}(k_K^2)\approx v_{\mathrm{F}}\hbar|k|,
\end{equation}
where we have set $E(k_K) = 0$. Note that the above expansion ignores
both the length scale of the graphene lattice constant $a=1.4$
\AA\ and the preferred directions of the lattice due to the discrete
lattice symmetry. In this low-$k$ limit, the Hamiltonian,
equation (\ref{H_Graph_TB}), can be approximated by the Dirac Hamiltonian,
\begin{equation}\label{Eq:Dirac}
 H=v_F\left(p_x\sigma_x\otimes\tau_0+p_y\sigma_y\otimes\tau_z\right)
\end{equation}
with $\vec{\sigma}$ and $\vec{\tau}$ being the Pauli spin matrices
acting on the pseudo-spin and valley degrees of freedom. Analytic
solutions for an infinitely extended graphene sheet described by
equation ({\ref{Eq:Dirac}) yields plane waves $\ket k$ where the angle
of the $k$ vector $\theta_k$,
\begin{equation}\label{Eq:Thetak}
\theta_k = \tan^{-1}(k_y / k_x),
\end{equation} 
 connects relative amplitudes on the $A$ and $B$ sublattice \cite{guinea_review},
\begin{equation}\label{pseudospin}
\ket{\mathbf{k}}= e^{i \mathbf{k\cdot r}}\left(\ket{A} + e^{i \theta_k}\ket{B}\right)/\sqrt 2.
\end{equation} 
Consequently, the pseudo-spin projection along the direction of
propagation, the ``helicity'',
\begin{equation}
\hat{h} = (\vecrep{\sigma}\;\cdot\;\vecrep k)/\left|k\right|\label{eq:hel}
\end{equation}
is conserved reflecting the chiral symmetry of the ideal graphene
sheet in the low-$k$ limit. The additional degeneracy of \emph{two}
non-equivalent cones (``valleys'') at the $K$ and $K'$ points in the
reciprocal lattice allows to formally represent the low-energy band
structure near $E=0$ in terms of Dirac-like four-spinors
$\ket\psi=(\psi_A^{K},\psi_B^{K}, \psi_A^{K'},\psi_B^{K'})$ with
amplitudes for the $A-B$ sublattice in real space and \KKp\ in
reciprocal space. The sign of $\theta_k$ [equation (\ref{Eq:Thetak})] is
reversed upon transition from $K$ to $K'$. (Note that physical spin is
not included in the present analysis.)

This Dirac-like picture may serve as valuable starting point for the
analysis of finite-size and edge effects on transport incorporated
within the tight-binding Hamiltonian [equation
  (\ref{H_Graph_TB})]. Chirality [equation (\ref{eq:hel})] is preserved in
the presence of slowly (on the scale of the C-C bondlength) varying
perturbations, suppressing backscattering \cite{guinea_review}
\begin{equation}
P(\vecrep{k}\rightarrow\vecrep{k}') =\left|\bra{\vecrep
  k'}V\ket{\vecrep k}\right|^2 \propto \cos^2[(\theta_{\vecrep k} -
  \theta'_{\vecrep k'})/2].\label{proptozero}
\end{equation}
Conversely, rough edges may break chirality resulting in
non-vanishing backscattering on the same cone and, at the same time,
in coupling of the $K$ and $K'$ cones. In the following, we will
investigate in detail the effects of short-range defects \cite{Ugeda}
(e.g. edges) and deviations from the continuum Dirac picture on the
transport properties of rough-edged graphene nanoribbons.

\section{Numerical Method}\label{sec:tech}

\begin{figure}
\hbox{}\hfill\epsfig{file=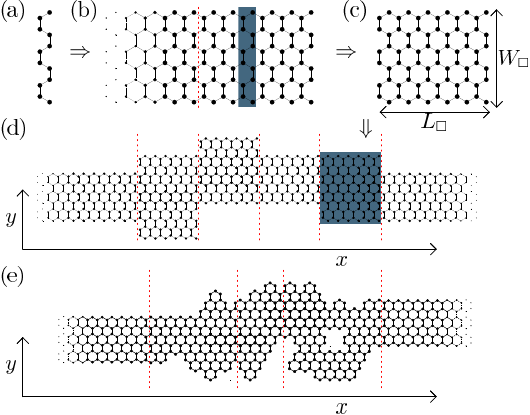, width = 8cm}\hfill\hbox{}
\caption{Assembling a rough-edged nanoribbon by combining several
  modules: (a) a chain of carbon atoms in transverse
  direction ($y$) is periodically repeated to yield (b) a
  half-infinite graphene ribbon along the $x$ direction. (c) Using
  Dyson equations [see equation (\ref{eq:dyson})], a rectangular
  region can be separated from the half-infinite ribbon. (d) A
  rough-edged ribbon can now be assembled by randomly combining
  rectangles of variable length $L_\Box$ and width $W_\Box$. (e) For
  an arbitrarily shaped graphene scattering structure, the modular
  approach can still be used, although the Green's function of each
  module has to be calculated by direct inversion.}
\label{fig:graph}
\end{figure}

For the numerical treatment of finite-size graphene flakes and ribbons
a number of simulation algorithms have meanwhile been proposed
\cite{MacKinnon2,bara,Lewenkopf,Heinzel,tra07,cres,meta,neop,sviz,
  Roche}.  We use in the following an extension of the
Modular Recursive Green`s function Method (MRGM)
\cite{PRB62RTWTB,Stephan, Rott06} applied to the third-order TB
Hamiltonian [equation (\ref{H_Graph_TB})]. The key idea of the MRGM is
to break down a large device into independent smaller modules, each of
which can be computed efficiently [see figure \ref{fig:graph}]. The
Green's functions $G_\Box$ of the different modules with width $W$ and
length $L$ are then combined to the desired device geometry using a
small number of Dyson equations. In this Article, we
  introduce an efficient method to calculate the $G_\Box$: the
  associated numerical effort becomes independent of module length
  $L$. The algorithm involves the somewhat counter-intuitive steps to
  first calculate infinitely and semi-infinitely extended ribbons,
  i.e., modules of width $W_{\Box}$ but $L=\infty$, from which 
  rectangular modules of finite length $L_\Box$ are ``cut out'' as needed by
  applying the Dyson equation ``in reverse''. The obvious advantage of
  this approach is that the computational effort becomes independent
  of $L_{\Box}$. This approach is particularly advantageous for the
simulation of weakly disordered graphene nano-ribbons: for weak
disorder the spacing between individual defects both in the bulk and
at the edges of the ribbon is large as compared to the lattice
spacing. We can thus simulate the region between neighbouring defects
by a single graphene module with perfect boundaries and place adjacent
defects at the module boundaries [see red dotted lines in figure
  \ref{fig:graph}(d)]. This efficient calculation of the
  extended rectangular modules in between two defects is key to
  simulating large devices with length $L_{\mathrm{total}}$ up to
  several micrometers (or $\approx 10^4$ hexagons in one direction).

As a prototypical example, we build up an infinitely long nanoribbon
with ideal zigzag boundaries (along the $\hat x$ direction) by
periodic repetition of a chain of carbon atoms of width $W$ in transverse
($\hat y$) direction [see figure \ref{fig:graph}(a)]. Other geometries
and boundaries can be treated analogously. The Hamiltonian $H$ of the
ribbon can thus be decomposed into a matrix $H_0$ describing the
Hamiltonian of the vertical chain, and the coupling matrix $H_I$
describing the connection between two adjacent chains \cite{Sanvito},
\begin{equation}
  H = \left(\begin{array}{ccccc}
\ddots&&&&\\
&H_0&H_I&0&\\
&H_I^\dagger&H_0&H_I&\\
&0&H_I^\dagger&H_0&\\
&&&&\ddots\\
\end{array}\right)\label{eq:H}.
\end{equation}
The solution of the Schr\"odinger equation for the infinite ribbon can be written 
in terms of an ansatz for a Bloch wave
\begin{equation}\label{eqketpsi}
\ket{\psi} = \sum_n e^{ik_nx}\ket{\chi_n}
\end{equation}
with $\ket{\chi_n}$ the transverse eigenfunction. For expanding the
Green's function in terms of $\chi_n$ we need a complete set of
transverse eigenfunctions including all evanescent modes
in the sum [equation (\ref{eqketpsi})] and the subsequent
equations. The resulting generalized eigenvalue problem for
$e^{ik\Delta x}$ and $\ket{\chi_n}$ gives $n$ left (right)-moving
states $\ket{\chi_j}$ ($\ket{\chi_{\jj}}$), with corresponding
momentum $k_j$ ($k_{\jj}$) in $x$ direction. In the following, we
introduce the shorthand notation $D_j(x) = \ket{\chi_j}e^{i k_j
  x}\bra{\chi_j}$ ($D_\jj(x) = \ket{\chi_\jj}e^{ik_\jj
  x}\bra{\chi_\jj}$) for the projections onto the right (left) moving
Bloch states. From the Bloch states the Green's function of the
infinite ribbon follows as \cite{Sanvito}
\begin{equation}
G^{\infty}(x,x') = \left\{
\begin{array}{rr}
\sum_{j=1}^N D_j(x-x') V^{-1},&x\ge x'\\
\vspace{-1mm}\\
\sum_{j=1}^N D_\jj(x-x') V^{-1},&x\ge x'\\
\end{array}
\right. , \label{eq:Ginf}
\end{equation}
with the hopping matrix 
\begin{equation}
V=\sum_{j=1}^N H_I^\dagger \left[ D_j(-\Delta x) - D_\jj(-\Delta x)\right].
\end{equation}
The Green's function of the half-infinite ribbon $G_{\mathrm{L}} (G_{\mathrm{R}})$
extending from $x_0$ to $-\infty$ (or $+\infty)$ can be written as 
\begin{equation}
G_{\mathrm{R,L}}(x,x') = G^{\infty}(x,x') + G^0_{\mathrm{R,L}}(x,x')
\label{eq:G0a}
\end{equation} 
with
\begin{eqnarray}
G_L^0(x,x') &=& \sum_j D_\jj(x-x_0) D_j(x_0-x') V^{-1}, \\
\label{eq:G0b}
G_R^0(x,x') &=& \sum_j D_j(x-x_0) D_\jj(x_0-x') V^{-1},
\label{eq:G0c}
\end{eqnarray}
satisfying the boundary conditions $G_{R,L} (x,x') = 0$ for all $x$ or
$x'$ located at the end of the half-infinite ribbon ($x,x'=x_0$).  For
an intuitive interpretation of equation (12) consider a disturbance
from a point source at $x'$. It reaches $x$ by two paths: the direct
propagation from $x'$ to $x$, given by the Green's function of the
infinite ribbon, and the propagation from $x$ to $x_0$ [given by
  $D_\jj(x-x_0)$ in equation (\ref{eq:G0b})], where the wave is
reflected at the end of the ribbon and then propagates from $x_0$ to
$x'$ [given by $D_j(x_0-x')$ in equation (\ref{eq:G0c})].  Note that
the numerical effort to calculate $G^\infty$ and $G_{\mathrm{R,L}}$ is
controlled by the transverse width of the ribbon $W_\Box$ and the
number of transverse modes $\ket{\chi_n}$ to be included while the
$x$-dependence is given analytically. This scaling behaviour is key to
calculate $G_{\Box}$ for the rectangular ribbon of arbitrary length
$L_{\Box}$ by solving the Dyson equation
\begin{equation}
  G_{L,R} = G_\Box + G_\Box H_I G_{L,R} \label{eq:dyson}
\end{equation}
in reverse for $G_\Box$ instead of for $G_{L,R}$. Consequently, the
numerical effort to calculate $G_\Box$ is independent of $L_\Box$. In
the final step, a rough-edged nanoribbon can now be assembled by
successively joining rectangular ribbons $G_\Box^{(i)}$ of varying
length $L_\Box$ and width $W_\Box$ (average width $\overline W =
60$nm) using the Dyson equation in forward direction,
\begin{equation}
G = G_\Box^{(i)}+G_\Box^{(i)}H_IG.\label{dyson:forward}
\end{equation}
In our simulations, we treat ribbon lengths of several micrometers,
and average over 100 random realizations $\xi$ of edge roughness to
eliminate non-generic features of particular ribbon configurations. To
assemble such very long disordered ribbons, we start with a set of
$N_B$ different modules ${M_1,\ldots,M_{N_B}}$ and combine them to
obtain a larger module $M_{N_B +1}$. We connect the calculated modules
in a random permutation $\mathcal{P}$,
$M_{N_B+1}=\mathcal{P}(M_1+\ldots+M_{N_B})$ (e.g., for $N_B =5$, $M_6
= M_3+M_1+M_5+M_4+M_2$). This procedure is repeated iteratively [i.e.,
  $M_7 = \mathcal{P}(M_2 + \ldots + M_6)$, formally equivalent to the
  composition rule of a Fibonacci sequence], creating an exponentially
growing, pseudo-random sequence of modules. The interfaces between
modules that include the disorder are randomly determined at each
iteration step to avoid periodic repetition.

If a more general shape of the scattering geometry is desired
[e.g., for a non-separable disorder potential or curved boundaries,
see figure \ref{fig:graph}(e)], the partitioning into modules and the
subsequent efficient buildup of long structures is still readily
possible.  Only the first step of our algorithm has to be
modified. The Green's function of individual modules is directly
calculated by inversion, i.e., $G=(E-H)^{-1}$, using, e.g., a
parallelized sparse-matrix solver \cite{MUMPS}. Subsequent application
of Dyson equations allows to assemble complex
scattering geometries.

We find that the computing time $\tau$ of our approach scales as $\sim
W^3$ due to the cubic dependence of the eigenvalue problem and of the
matrix multiplications. $\tau$ scales linearly with the number of
building blocks $N_B$ used to set up the geometry, and logarithmically
with $L/L_\Box$. Numerically, we find for $\tau$,
\begin{equation}\label{eq:numtime}
\tau [ms] \sim a \; N_B\,\ln\left(\frac{L}{L_\Box}\right)
\cdot W^3,
\end{equation}
with $W$ given in nm. We have determined a prefactor $a \approx 5$
from calculating the full scattering problem averaged over 100
configurations, when computing on 3 AMD Opteron processors (24 cores)
at 2.2 GHz. Clearly, the prefactor strongly depends on the details of
the employed hardware (i.e., network speed, cache size, compilation
flags, etc.), while the scaling (\ref{eq:numtime}) does not.

We note that the application of the algorithm presented here is not
restricted to graphene nanostructures: any modular scattering system,
which is build from modules along the lines of figure 2 (a)-(c) can 
be treated analogously. Possible applications include acoustic
cavities, conventional semiconductors, topological insulators, or
neutron scattering devices.

\section{Results}\label{sec:rec}

\begin{figure}
  \hbox{}\hfill\epsfig{file=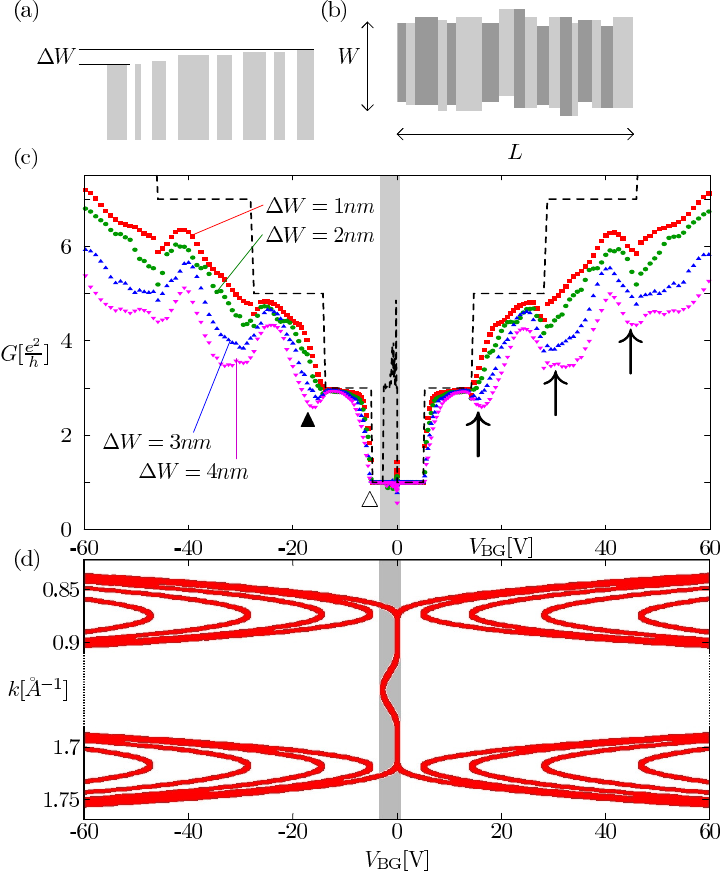,width=10cm}\hfill\hbox{}
\caption{{(a) Eight building blocks of different length $l \pm \Delta
    l = 3 \pm 2 nm$ and height in a range $W \pm \Delta W/2$ were used
    to assemble (b) a rough-edged graphene nanoribbon (different
    shades of grey for clarity). (c) Ensemble-averaged conductance $G$
    of 40nm wide graphene ribbons of length $L=100nm$ with different
    amplitude of edge roughness $\Delta W$ as a function of back-gate
    voltage $V_{\mathrm{BG}}$. The conductance of a ribbon with
    perfect zigzag boundaries is shown as dashed black line. Arrows
    ($\uparrow$) mark dips in the conductance (see text). The shaded
    area highlights the voltage interval of increased conductance in
    the ideal ribbon due to states localized at the zigzag edge (see
    text). The solid $\blacktriangle$ [open $\vartriangle$] triangle
    marks the back gate voltage of individual scattering states
    displayed in figure \ref{fig:graph:pic} (a) [(b)]. (d) Dispersion
    relation $k[E(V)]$ of an ideal 40 nm wide graphene zigzag ribbon,
    enlarged around the $K$ and $K'$ points.}}
\label{fig_bias}
\end{figure}

\begin{figure}
  \hbox{}\hfill\epsfig{file=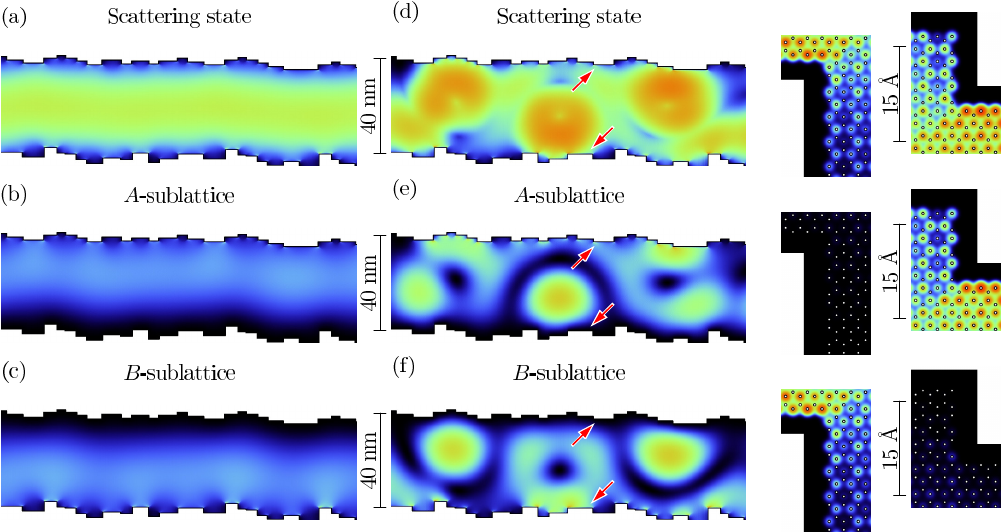,width=12cm}\hfill\hbox{}
\caption{{Scattering states of rough-edged graphene nanoribbons at
    selected back gate voltage: (a)-(c) $V_{BG}=-5$ V [corresponding to
    open triangle in figure \ref{fig_bias}(c)], (d)-(f)
    $V_{BG}=-15$ V [solid triangle in figure
    \ref{fig_bias} (c)]. Panels (a), (d) show the entire scattering
    wavefunction, while panels (b,e) [(c,f)] feature projections onto
    the A [B] sublattice respectively. Frames to the right show zoom-ins
    of wave function enhancements at upper (lower) corners marked by
    red arrows in (d)-(f), the positions of the carbon atoms are
    marked by white dots as guide to the eye.}}
\label{fig:graph:pic}
\end{figure}

\subsection{Transport coefficients}

For conventional semiconductor heterostructures (e.g., quantum dots
made of GaAs-AlGaAs), confinement is usually achieved by electrostatic
gates resulting in smooth dot boundaries. Such confinement is not
realizable for graphene due to its gapless band structure. While
several theoretical concepts for opening a band gap have been
proposed, the majority of experiments have achieved confinement by
patterning of graphene nanodevices with oxygen plasma etching, chemical
vapor deposition, specially prepared SiC substrates \cite{deHeer}, or chemical
etching. These techniques, however, do not result in well-defined
armchair or zigzag edges but in a rough-edge pattern featuring
armchair and zigzag elements as well as adsorbates at the dangling
carbon bonds \cite{pon08, sta08aa, sta08ab, KimNature} leading to an
irregular edge structure. Edge effects can thus be expected to
strongly influence the properties of graphene nanodevices.

We simulate the influence of edge scattering on transport through
graphene nanoribbons by randomly varying the widths of the rectangular
modules which build up the ribbon in the range $W = 40\pm$1 nm
[see figure \ref{fig_bias}(a)]. Numerical tests show that a random
sequence of $N_B = 5$ different module widths represents a good
compromise between a high degree of randomness and limited
computational effort. In order to suppress correlations in the
$x$-dependence of the roughness, the length of each rectangular module
is chosen at random in the range of 0.24nm (one unit cell) to 10nm. We
then use the above Fibonacci-like procedure to assemble a scattering
geometry [see figure \ref{fig_bias}(b)] of up to several $\mu$m in
length. Finally, all modules are connected to two ideal half-infinite
graphene waveguides. We average over 100 realizations $\xi$ of
nanoribbons to eliminate non-generic features of particular ribbon
configurations.

In addition to the quantization steps due to transverse confinement
[dashed black line in figure \ref{fig_bias}(c)], a graphene nanoribbon
with a perfect zigzag boundary of fixed width $W$ features edge states
with finite dispersion [figure \ref{fig_bias}(d)] since the coupling
between the outermost carbon atoms is non-zero
\cite{dassarma_review}. Consequently, the edge states of an ideal
nanoribbon give rise to a peak in conductance $G$ just below the Dirac
point [shaded area in figure \ref{fig_bias}(c,d)]. In contrast to
first-nearest-neighbor tight binding, and in line with the full
ab-initio bandstructure and experiment \cite{dassarma_review}, our
third nearest neighbor approach accounts for the breaking of
electron-hole symmetry. Conductance is thus only approximately
symmetric relative to $E=0$.

In the presence of edge disorder, $G$ undergoes several pronounced
changes [figure \ref{fig_bias}(c)]: overall, $G$ decreases with
increasing distance in energy from the Dirac point relative to the
ideal ribbon. The quantization steps due to the transverse confinement
are strongly suppressed. Moreover, the edge disorder completely
removes the sharp conductance peak attributed to edge states, as they
are no longer conducting but become localized parallel to the ribbon
\cite{LibPRB08}, i.e.~along the direction of transport. Consequently,
corresponding signatures are difficult to observe in transport
measurements of realistic samples. Scanning tunneling spectroscopy
provides an alternative approach: peaks in the local density of states
at energies slightly below the Dirac point have been recently, indeed,
observed in STS experiments \cite{niimi}.

In the limit where quantization steps due to the transverse
confinement are strongly suppressed [figure \ref{fig_bias}(c)]
pronounced broad dips in transmission (see arrows in figure
\ref{fig_bias}) replace the original steps in conductance. This
counter-intuitive \emph{reduction} of transmission with
\emph{increasing} energy in the vicinity of steps can be qualitatively understood by
considering Fermi's golden rule for the scattering of mode $\ket{n k}$
into mode $\ket{n' k'}$ \cite{Ihnatsenka,LibKirchberg2011},
\begin{equation}\label{eqFGR}
\Gamma(E) \propto \sum_{n'} \left|\left<nk|H|n'k'\right>\right|^2 \rho_{n'}(E).
\end{equation}
Two trends contribute to this effect: firstly, strong fluctuations of
the ribbon width broaden the DOS $\rho_{n'}(E)$ and smoothen the
steps. Secondly, as the ribbon locally narrows, backscattering via
scattering into evanescent modes is enhanced. This occurs
preferentially for energies close to the opening of a new mode and
results in a reduction in transmission causing the dips.  

It is instructive to compare the present results to calculations for
edge- and bulk- disordered semiconductor nanowires featuring a
parabolic dispersion relation in the long-wavelength (continuum)
limit. While a reduction of quantization steps by disorder is observed
for edge-disordered semiconductor ribbons \cite{Nikolic1994}
resembling the present results, the prominent transmission dips
observed for graphene (arrows in figure 3) appear not to be present in
such a system (compare with figure 2 in \cite{Nikolic1994}). However,
dips have been found in other disordered semiconductor nanostructures
that are associated with resonances supported by attractive (bulk)
disorder potentials \cite{Bagwell}. The distance in energy between
these resonances and the quantization step (i.e.~the subband minimum)
corresponds to the binding energy of the (quasi) bound state
\cite{Bagwell}. In the present case of graphene with rough edges, the
enhancement of the local density of states near the Dirac point
resulting from localized states at the edges is well known
\cite{Heinzel}. Their statistical weight has been found to be much
higher for graphene than for conventional nanostructures
\cite{LibPRB08}. In the absence of a local attractive potential, these
localized states could take on the role of resonances: for edge
structures similar to the ones we investigate, the resonance energies
$E_i$ are statistically distributed in the range $E_i \in [-80,0]$ meV
\cite{LibPRB08}, and could, possibly, give rise to the observed
broad dips.

\begin{figure*}
\hbox{}\hfill\epsfig{file=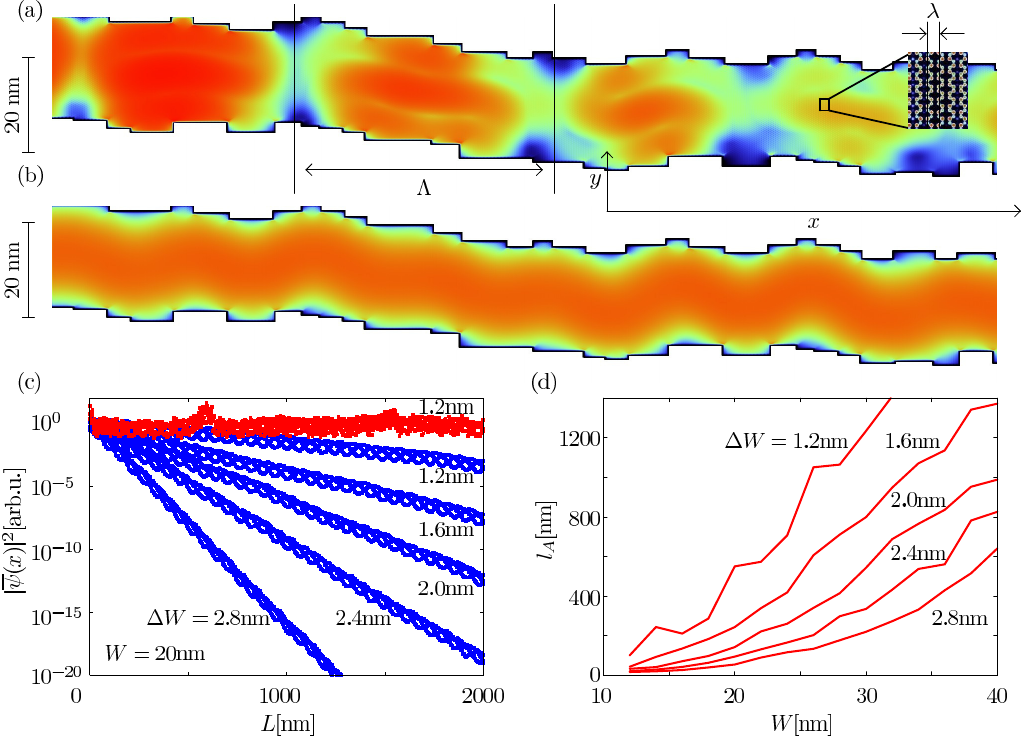,width=15cm}\hfill\hbox{}
\caption{{(a) Anderson-localized scattering state shown for a section
    (1100-1200nm) of a ribbon with total length $L=3\mu m$. (b)
    Conducting scattering state [as in figure \ref{fig:graph:pic}(a)]
    shown for the same section as in (a). (c) Longitudinal dependence
    $\left|\bar{\psi}(x)\right|^2$ of the conducting state [red, see
      (b)], and localized states [blue]. The latter are averaged over
    100 Anderson-localized states for average ribbon width $W = 20$nm
    and different edge roughness amplitude $\Delta W$ (see insets).
    (d) Localization length $l_A$ as function of ribbon width $W$ at
    energy $E=0.2eV$ for the five different values of edge roughness
    $\Delta W$ as in (c).  }}
\label{fig:Anderson}
\end{figure*}

\subsection{Localized scattering states at edges}

To gain a deeper understanding of the
transport characteristics of edge-disordered graphene ribbons, we now
analyze also individual scattering states. We find
that states with energies where the conductance is only weakly
perturbed by edge disorder [e.g.~open triangle in figure
  \ref{fig_bias}(c)] feature a low amplitude at the edges [see figure
  \ref{fig:graph:pic}(a)]. The overall probability density of these
scattering states remains concentrated near the center of the ribbon
[see figure \ref{fig:graph:pic}(a)] and is
therefore only slightly affected by edge disorder. When only a single
mode is open in the leads (at energies close to the Dirac point),
modes located at $K$ and $K'$ in momentum space are not coupled in a
zigzag graphene nanoribbon \cite{guinea_review}. Only one cone
contributes to transport in each direction [see figure
  \ref{fig_bias}(d)]. This imbalance in the number of left- and
right-moving channels on each cone is a special property of zigzag
graphene nanoribbons \cite{Wakaba}, similar to the band structure of
topological insulators. Backscattering is only possible in this energy
window by inter-valley scattering at the rough edges. Since we observe
a nearly perfectly conducting channel [figure
  \ref{fig:Anderson}(b),(c)], inter-valley scattering that requires
momentum transfers of the order $\left|K-K'\right|$ is obviously
suppressed at low energies \cite{XiongA}.

By contrast, scattering states at energies where transmission (and
conductance) is considerably reduced [solid triangle in figure
  \ref{fig_bias}(c) and figure \ref{fig:graph:pic}(b)] feature a
strong enhancement of their wavefunction near corners of the edges
originating from one sub-lattice only [see arrows in figure
  \ref{fig:graph:pic}(e, f)].  Projections onto the $A$ and $B$
sublattices [figure \ref{fig:graph:pic} (e, f)] show pronounced
differences reflecting the violation of pseudo-spin conservation
[equation (\ref{proptozero})]. We find enhancements of the $A$ ($B$)
sub-lattice scattering wave function at the upper (lower) edges of the
ribbon, i.e., at those edges where the outermost carbon atom is of
type $A$ ($B$) [see zoom-ins in figure \ref{fig:graph:pic}(d)-(f)], in
line with a strong enhancement of the local DOS near rough edges
\cite{Heinzel,guinea_review, dassarma_review}.  However, the
pronounced differences in the wavefunction patterns near the center of
the ribbon are not accounted for only by localized edge states since
their decay length into the ribbon interior is much smaller than the
ribbon width. We therefore attribute the dramatic drop in conductance
to pronounced intra-valley and inter-valley backscattering at the edge
corners, since the suppression of backscattering associated with the
conservation of pseudo-spin [equation (\ref{proptozero})] no longer
holds.

As reported in earlier work on edge disorder in rough-edged graphene
nanoribbons, transmission is strongly suppressed close to the Dirac
point, leading to the formation of a transport gap \cite{ Lewenkopf,
  Heinzel}. Atomic-scale defects on the edges of wide ribbons may lead
to exponential (i.e., Anderson) localization due to destructive
interference \cite{Heinzel, XiongA, Nikolic1994}. We use our modular
approach to calculate scattering states on mesoscopic length scales
[ribbon length $L=2\mu$m, see figure \ref{fig:Anderson}(a, b)]. By
averaging over many realizations of edge disorder, we can thus
explicitly probe for exponential localization and determine
the localization length. Looking at the longitudinal dependence of the
scattering state,
\begin{equation}
  |\bar{\psi}(x)|^2 = \int_0^W |\psi(x,y)|^2 \mathrm dy\label{norm:proj},
\end{equation}
we observe an exponential decay over up to 10 orders of magnitude 
[figure \ref{fig:Anderson}(c)]. Fitting to the functional form
$\left|\bar{\psi}(x)\right|^2\propto \exp(-x/l_A)$ we can numerically
extract the localization length $l_A$. We find $l_A$ to scale as
$l_A\approx \alpha W/\Delta W$, i.e., $l_A$ increases linearly with
ribbon width and is inversely proportional to the disorder amplitude
$\Delta W$ [figure \ref{fig:Anderson}(d)]. The localization length $l_A$
is found to increase with increasing distance (in energy or in $k$)
from the Dirac point (not shown), as suggested by the disorder-induced
formation of a transport gap \cite{Lewenkopf, Heinzel, Roche}. 

Superimposed on the exponential decay are oscillations on two shorter
length scales: (i) a short beating period of $\lambda = 0.7 nm$ due to
interference between the $K$ and $K'$ cones \cite{LibPRB08}
[$\lambda$ in figure \ref{fig:Anderson}(a)] and (ii) a much slower
variation with the length scale $\Lambda \approx 30nm$ [$\Lambda$ in
  figure \ref{fig:Anderson}(a)] which corresponds to the wavelength
$\Lambda=2\pi/k$ associated with the linear dispersion relation
$E=v_{\mathrm F} \hbar k$, i.e., the distance in $k$ space from the
$K$ point.

For comparison we also plot for the nearly perfectly conducting
channel its wavefunction and its projection according to equation
(\ref{norm:proj}) [figure \ref{fig:Anderson}(b),(c)]. If the incoming
scattering wave couples to the near-perfectly conducting channel,
this contribution will be dominant after a certain ribbon length as
all other contributions quickly die out. The oscillations due to
$K-K'$ interferences (i) are also present for this conducting state,
though at reduced amplitude. While we observe Anderson localization
for incoming scattering states at energies where more than one mode is
open per cone, near-perfect conduction \cite{Wakaba, XiongA} appears
to be confined to the topologically insulating part of the band
structure. We expect these states to have a localization length that
exceeds the dimension of our structure, if it is, at all, finite.

\subsection{Variations of edge roughness}

To investigate to what extent the above results depend on our
particular choice of rectangular edge roughness we generalize our
approach to include randomly jagged edges. We combine graphene segments
featuring horizontal zigzag edges with segments featuring a boundary
profile tilted by an angle $\beta$ with respect to the horizontal
zigzag direction [see insets in figure \ref{fig:jagged}]. As outlined
in section \ref{sec:tech}, we calculate a set of modules (we use
modules with length $L\in[20,40]$nm) by direct inversion of a
finite-sized Hamiltonian, and combine these modules to efficiently
generate very long structures (total length $>1\mu$m).

\begin{figure}
\hbox{}\hfill\epsfig{file=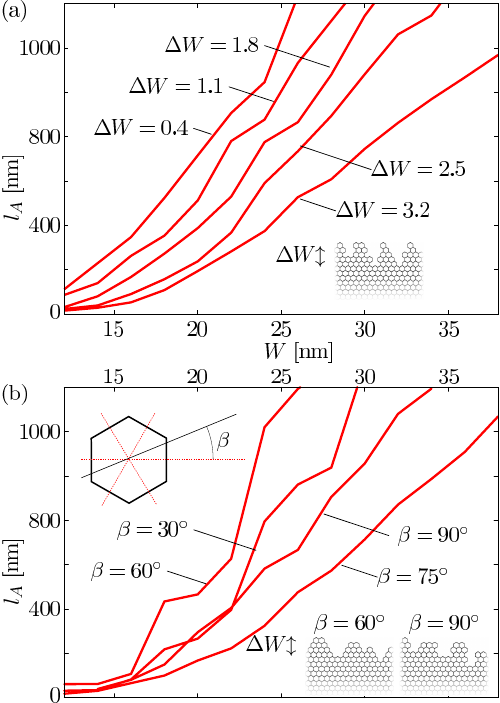,width=7cm}\hfill\hbox{}
\caption{{(a) Localization length $l_A$ as a function of ribbon width $W$
    for different values of edge roughness amplitude $\Delta W$ (see inset). Each
    curve is averaged over 100 disorder configurations, featuring
    random edge directions. (b) Same as (a) for fixed $\Delta W=2.0nm$, for
    different edge roughness configurations: the parameter $\beta$
    labels the angle (in degrees) between straight-line segments and
    the horizontal zigzag direction of the graphene lattice (see top left
    inset), resulting in different roughness configurations (see
    bottom right inset).}}
\label{fig:jagged}
\end{figure}

Qualitatively, we find the same Anderson localization behaviour [see
  figure \ref{fig:jagged}(a)] as a function of ribbon width $W$ and
roughness amplitude $\Delta W$ (for fixed $\beta$) as in the case with
rectangular modules. However, unlike the case of
  free-particle dispersion \cite{Nikolic1994}, graphene nanostructures
  feature an interesting interplay between lattice orientation and
  surface roughness. As this interplay is determined by the
  alignment angle $\beta$ between the lattice orientation and the
  roughness, we can explicitly study its influence on transmission
  through the ribbon. We observe, indeed, that the value
of the localization length strongly depends on the shape of the
boundary with respect to the discrete lattice: edges consisting of
randomly concatenated zigzag-edges only (i.e., with $\beta =
60^\circ$) show substantially longer localization lengths than edges
formed by an even mixture of zigzag and armchair edges [i.e., with
  $\beta=75^\circ$].  The dependence of the localization length on
$\beta$ can be understood in terms of the length of undisturbed zigzag
(or armchair) edges: cutting close to a symmetry plane of the lattice
(i.e.~$60^\circ$ or $30^\circ$) results in comparatively longer
segments of zigzag (or armchair) edges. By contrast, a cut at
$75^\circ$ yields an irregular sequence of very short segments of
armchair and zigzag boundaries and thus strongly breaks the
translation symmetry of a clean zigzag (or armchair) edge.  As an
aside we note that the data presented in the previous subsection
includes a variation in ribbon direction, i.e., the ribbons are not
perfectly straight, notably in figure \ref{fig:Anderson}(a, b). This
amounts to an effective increase of edge roughness. As a result, the
localization length is further decreased in that case [compare figure
  \ref{fig:Anderson}(d) with figure \ref{fig:jagged}(b) for
  $\beta=90^\circ$].

The observation of the relative change in resistance as a function of
$\beta$, i.e., of the angle between the graphene lattice and the
atomic-scale edge, might have implications for experiments. Measuring
the atomic-scale roughness is difficult requiring an STM
setup. Measuring localization length for different ribbon widths might
provide an alternative probe for the atomic-scale edge
roughness. Conversely, our results could be tested by
  comparing transport measurements for nanoribbons fabricated with
  different methods (i.e.~etching, growth on Si-C
  substrates \cite{deHeer}, unzipping of graphene nanotubes
  \cite{Crommie}) resulting in (known) different edge
  characteristics.

\begin{figure*}
\hbox{}\hfill\epsfig{file=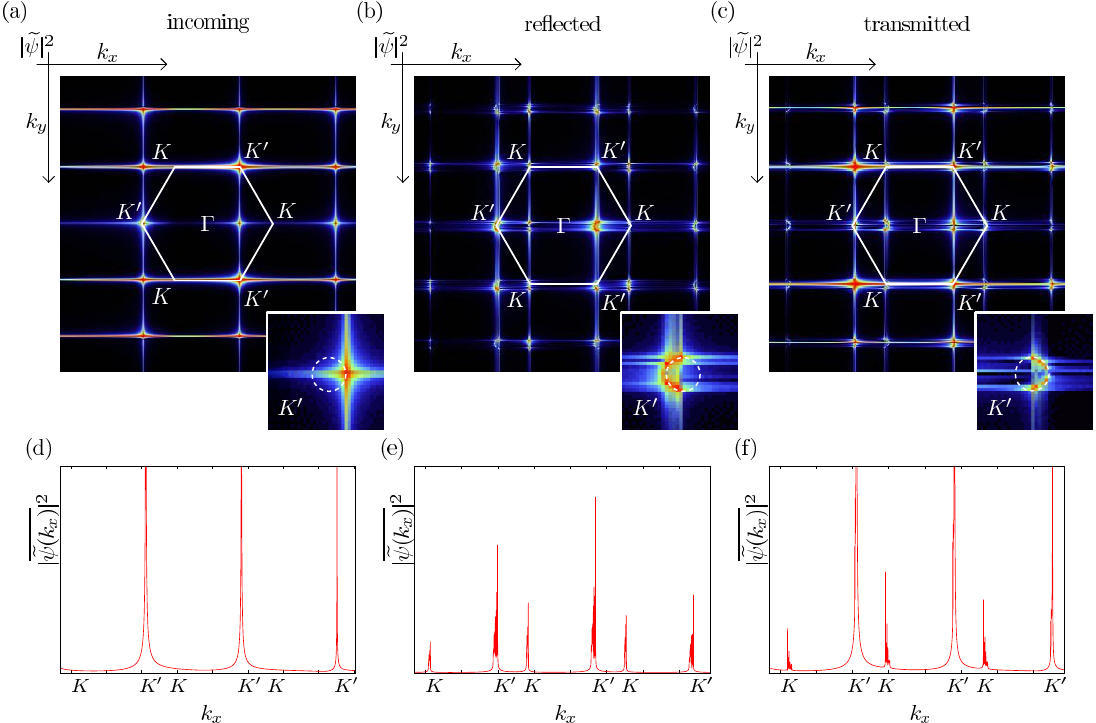,width=15cm}\hfill\hbox{}
\caption{{Two-dimensional Fourier transform $|\widetilde\psi(\mathbf
    k)|^2$ [top row], and longitudinal dependence
    $|\widetilde{\overline{\psi}}(k_x)|^2$ [bottom row, equation
      (\ref{four:projx})] of the incoming [(a,d), left column],
    reflected [(b,e), center column] and transmitted [(c,f), right
      column] part of the scattering state in the waveguides. Due to
    the finite size of the numerically evaluated scattering state, the
    Fourier transform features a grid of thin horizontal and vertical
    lines. The insets show an enlarged view of the $K'$ point [a
      dashed white circle is inserted as guide to the eye]. The first
    Brillouin zone of the reciprocal lattice is shown as white
    hexagon.}}
\label{fig:graph:four}
\end{figure*}

\subsection{Fourier analysis of channel states}

We explore now the interplay between short-range defects in real space
and the absence (presence) of $K-K'$ inter-cone scattering in
$k$-space. For this purpose we analyze the Fourier transforms of the
asymptotic scattering state in the semi-infinite entrance (exit)
waveguides. The Fourier transform is calculated as
 \begin{equation}\label{Fourier}
\widetilde\psi(\mathbf k)=\mathcal F[\psi(\mathbf r)] 
= \int_{\area}\mathrm d\mathbf r \;\psi(\mathbf r)\,
e^{i\,\mathbf{k\cdot r}},
\end{equation}
where we extend the integral over a finite area $\area$ in the
asymptotic region of the waveguide, i.e., far away from the scattering
region. Three different classes of asymptotic scattering states need to be
considered: (i) the incoming Bloch states propagating in $x$-direction
with wavenumber $k_n$,
\begin{equation}
\psi_n(\mathbf r) = e^{ik_{xn}x}\chi_n(y),
\end{equation}
where $\chi_n(y)$ represents the transverse eigenfunction of mode $n$
of the semi-infinite nanoribbon, (ii) the waves transmitted through the
disordered region with transmission amplitude $t_{mn}$, and (iii) the
reflected waves with reflection amplitude $r_{nm}$. The corresponding Fourier
components are given by 

\begin{eqnarray}
\widetilde\psi_n^T(\mathbf k)&=&\sum_m\mathcal F[t_{mn}\cdot\psi_m(\mathbf r)],\\
\widetilde\psi_n^R(\mathbf k)&=&\sum_m\mathcal F[r_{mn}\cdot\psi_m(\mathbf r)].
\label{Eq:BlochF}\end{eqnarray}

For a hexagonal lattice, the real and the reciprocal lattice are
rotated by 90 degrees with respect to each other \cite{Ashcraft}. The
first Brillouin zone for the ideal zigzag ribbon is thus given
by a hexagon resting on a side rather than on a tip [see white hexagon in
  figure \ref{fig:graph:four}(a)]. 
To better visualize
the enhancement of $\widetilde\psi(\mathbf k)$ near the $K$ or $K'$
points, we integrate $\widetilde\psi(\mathbf k)$ over the transverse
direction,
\begin{equation}
\left|\widetilde{\overline{\psi}}(k_x)\right|^2 = \int \mathrm dk_y |\widetilde\psi(\mathbf k)|^2.
\label{four:projx}
\end{equation}
In perfect zigzag ribbons, incoming modes feature non-vanishing
amplitudes either near the $K$ or the $K'$ point, i.e., there is no
coupling (scattering) between $K$ and $K'$. For the incoming Bloch
state the projected Fourier transform equation (\ref{four:projx})
features peaks at the $k_x$ values corresponding to $K'$ [see figure
  \ref{fig:graph:four}(d)]. The close-up of the peak in
$\widetilde\psi(\mathbf k)$ near the $K'$ point [inset of figure
  \ref{fig:graph:four}(a)] is structureless for the incoming Bloch
wave with fixed transverse quantum number $n$. The horizontal and
vertical lines are finite-size effects of the Fourier-transformed
sample. Likewise, the origin of the additional bright spots inside the
Brillouin zone is zone folding (the Brillouin zone of the ribbon is
smaller than the graphene Brillouin zone). The interesting physics, on
the other hand, is contained in finite amplitudes at both $K$ and $K'$
points of the scattered wave [see figure \ref{fig:graph:four}(b,c)]
which are induced by \KKp scattering at rough edges. The relative
strength of the integrated $K$ and $K'$ peaks [see figure
  \ref{fig:graph:four}(e,f)] is a direct measure for the amount of
inter-valley scattering.  Furthermore, we observe a pronounced fine
structure near the $K'$ and $K$ points: enhancement along a
half-circle forms around the Dirac points [see inset in figure
  \ref{fig:graph:four}(b,c)]. The surface of section of the
double-cone band structure of constant energy is approximately a
circle, the diameter of which is proportional to the energy. In the reflected
(transmitted) part of the wavefunction, we only see the left half
(right half) of this circle being populated, corresponding to negative
(positive) group velocities. Enhancement along the full semicircle is
due to inter-mode scattering $n\rightarrow m$ between transverse modes
within the same valley (intra-valley scattering). We can thus conclude
that pronounced intra-valley scattering at the rough edges distributes
the reflected (or transmitted) wave almost uniformly over the
energetically accessible half-circle of the band structure compatible
with their propagation direction. The Fourier transform thus allows us
to assess the amount of both inter-valley \KKp\ scattering (by the
relative amplitude around the $K$ and $K'$ points in the reciprocal
lattice) and inter-mode scattering by the angular distribution on the
half-circle of a single cone for the incoming and reflected (or
transmitted) states.

\section{Conclusions and Outlook}\label{Sec:sum}

We have presented a novel numerical approach to efficiently calculate
the Green's function for extended nanoribbons. Key is the build-up of
the ribbon by a random assembly of modules. Connecting these modules
by way of a Dyson equation allows us to calculate the transport
properties of long graphene ribbons. We find the conductance to be
suppressed by rough edges relative to that of the perfect ribbon. For
low energies, we observe near-perfectly conducting channels due to the
band structure of zigzag graphene nanoribbons. Quantization steps are
washed out and, in part, replaced by dips due to scattering into
evanescent modes \cite{Ihnatsenka2}, in contrast to
  edge-disordered semiconductor nanoribbons with free-particle dispersion
  \cite{Nikolic1994}. An analysis of individual scattering states in
both real space and Fourier space reveals pronounced \AB\ sublattice
asymmetries and \KKp\ scattering. We determined specific
  signatures of inter- and intra-valley scattering by Fourier
  transform spectroscopy of scattering states.  We also identified
  Anderson localized states for different disorder configurations,
  extending over several micrometers with an exponential decay
  spanning 10 orders of magnitude. The corresponding localization
length was calculated as a function of both the magnitude of edge
roughness, and its alignment with the graphene lattice. We find that
the latter plays a significant role in determining the localization
length hinting at the importance of correctly modeling microscopic
details of edge disorder beyond its amplitude and correlation
  length.

We conclude by pointing to possible future applications. While early
transport measurements were strongly affected by substrate
interactions resulting in puddles of electron and hole conductivity
due to bulk disorder, recent advances in the manufacturing of much
cleaner graphene nanostructures by growing on Si-C substrates
\cite{deHeer}, unzipping nanotubes to arrive at smooth-edged ribbons
\cite{Crommie} as well as suspended graphene \cite{vanWees} have
shifted the focus to edge disorder investigated in the present work.
Indeed, the measurement of size quantization plateaus has been
surprisingly elusive in graphene nanoribbons
\cite{Molitor,lin08,Kim2010,LewenkopfReview}, in qualitative agreement
with our present findings. Only recently, by prolonged annealing and
suspending graphene nanoconstrictions, first signatures of size
quantization could be found \cite{vanWees}. Our findings
regarding the sublattice sensitivity of the wavefunction (figure 3)
could be tested by STM scans of bound states in graphene nano-islands
\cite{Dinesh}: in these measurements strong enhancements of
wavefunction amplitudes were found on one sublattice,
resulting in trigonal patterns close to edges that are quite similar
to our numerical findings. Our simulations predict similar STM
patterns for scattering states in extended nanostructures. In
particular, such measurements could elucidate the precise nature of
the scattering mechanisms encountered at edges prepared wth different
techniques. Indeed, we expect defects affecting both sublattices, as
recently investigated \cite{Ugeda}, to exhibit different signatures
than e.g.~single vacancies. Finally, magnetic field effects allow for
an additional external parameter more easily tunable in experiments
than lattice geometries.  States associated with different K points
react differently to magnetic fields. This dependence might help to
disentangle contributions from different $K$-points in scattering
accessible by our Fourier analysis. We note that our algorithm can
be easily adapted to accomodate magnetic fields while retaining its
favorable scaling properties. Investigations in this direction are
currently under way.

\acknowledgements{We thank K.~Anson, L.~Chizhova, J.~G\"uttinger, and C.~Stampfer
for valuable discussions.  Support by the Austrian Science Foundation
(Grant No.~FWF-P17359), the Max Kade Foundation and the SFB 041-ViCoM
(FWF) is gratefully acknowledged. S.R. acknowledges support by the
Vienna Science and Technology Fund (WWTF) through Project No.
MA09-030 and by the Austrian Science Fund (FWF) through Project
No. P14 in the SFB IR-ON. Numerical calculations were performed on
the Vienna scientific cluster (VSC).}

\end{document}